\documentclass[aps,prd,twocolumn,nofootinbib]{revtex4-1}
\usepackage{graphicx}
\usepackage{amsmath}
\usepackage{graphicx}
\usepackage{amssymb}
\usepackage{color}

\def\figwidth{0.47\textwidth}
\def\hunit{\,\mathrm{km\,s^{-1}Mpc^{-1}}}
\def\be{\begin{equation}}
\def\ee{\end{equation}}
\def\bq{\begin{eqnarray}}
\def\eq{\end{eqnarray}}
\def\page{p_{\rm age}}
\def\aeff{a+2\log_{10}\frac{h}{0.7}}

\begin{document}

\title{Reconciling low and high redshift GRB luminosity correlations}

\author{Lu Huang}
\affiliation{School of Physics and Astronomy, Sun Yat-Sen University, 2 Daxue Road, Tangjia, Zhuhai, 519082, P.R.China}
\author{Zhiqi Huang$^*$}
\affiliation{School of Physics and Astronomy, Sun Yat-Sen University, 2 Daxue Road, Tangjia, Zhuhai, 519082, P.R.China}
\email{huangzhq25@mail.sysu.edu.cn}
\author{Xiaolin Luo}
\affiliation{School of Physics and Astronomy, Sun Yat-Sen University, 2 Daxue Road, Tangjia, Zhuhai, 519082, P.R.China}
\author{Xinbo He}
\affiliation{School of Physics and Astronomy, Sun Yat-Sen University, 2 Daxue Road, Tangjia, Zhuhai, 519082, P.R.China}
\author{Yuhong Fang}
\affiliation{School of Physics and Astronomy, Sun Yat-Sen University, 2 Daxue Road, Tangjia, Zhuhai, 519082, P.R.China}

\begin{abstract}

  The correlation between the peak spectra energy ($E_p$) and the equivalent isotropic energy ($E_{\rm iso}$) of long gamma-ray bursts (GRBs), the so-called Amati relation, is often used to constrain the high-redshift {Hubble diagram}. Assuming Lambda cold dark matter ($\Lambda$CDM) cosmology, Ref.~\cite{Wang:2017lng} found a $\gtrsim 3\sigma$ tension in the data-calibrated Amati coefficients between low- and high-redshift GRB samples. {To reduce the impact of fiducial cosmology, we use the Parameterization based on cosmic Age (PAge), an almost model-independent framework to trace the cosmological expansion history. We find that the low- and high-redshift tension in Amati coefficients stays almost the same for the broad class of models covered by PAge, indicating that the cosmological assumption is not the dominant driver of the redshift evolution of GRB luminosity correlation. Next, we analyze the selection effect due to flux limits in observations. We find} Amati relation evolves much more significantly across energy scales of $E_{\rm iso}$. {We debias the GRB data by selectively discarding samples to match low-$z$ and high-$z$ $E_{\rm iso}$ distributions. After debiasing, the Amati coefficients agree well between low-$z$ and high-$z$ data groups, whereas the evidence of $E_{\rm iso}$-dependence of Amati relation remains to be strong. Thus, the redshift evolution of GRB luminosity correlation can be fully interpreted as a selection bias, and does not imply cosmological evolution of GRBs.}

\end{abstract}

\maketitle

\section{Introduction}\label{intro}

Since the first discovery of the cosmic acceleration, Type Ia supernovae (SNe) have been employed as standard candles for the study of cosmic expansion and the nature of dark energy~\cite{Perlmutter:1998hx,perlmutter1999astrophys,riess1998observational,Scolnic:2017caz}. Due to the limited intrinsic luminosity and the extinction from the interstellar medium, {the maximum redshift of currently detectable SNe is about 2.5~\cite{Scolnic:2017caz}.} This at the first glance seems not to be a problem, as in the standard Lambda cold dark matter ($\Lambda$CDM) model, dark energy (cosmological constant $\Lambda$) has negligible contribution at high redshift $z\gtrsim 2$. However, distance indicators beyond $z\sim 2$ can be very useful for the purpose of testing dark energy models beyond $\Lambda$CDM. One of the attractive candidates is the long gamma-ray bursts (GRBs) that can reach up to $10^{48}$-$10^{53}\mathrm{erg}$ in a few seconds. These energetic explosions are bright enough to be detected up to redshift $z\sim10$~\cite{Piran:1999kx,Meszaros:2001vi,Meszaros:2006rc,Kumar:2014upa}. Thus, GRBs are often proposed as complementary tools to Type Ia supernova observations.  Due to the limited understanding of the central engine mechanism of explosions of GRBs, GRBs cannot be treated as distance indicators directly. Several correlations between GRB photometric and spectroscopic properties have been proposed to enable GRBs as quasi-standard candles~\cite{Amati:2002ny,Ghirlanda:2004fs,Amati:2008hq,Schaefer:2006pa,Capozziello:2008tc,Dainotti:2008vw,Bernardini:2012sm,Amati:2013sca,Wei:2013wza,Izzo:2015vya,Demianski:2016zxi,Demianski:2016dsa}. The most popular and the most investigated GRB luminosity correlation is the empirical Amati correlation between the rest-frame spectral peak energy $E_{p}$ and the bolometric isotropic-equivalent radiated energy $E_{\rm iso}$, given by a logarithm linear fitting
\begin{equation}
  \log_{10} \frac{E_{\rm iso}}{\mathrm{erg}} = a + b  \log_{10} \frac{E_{p}}{300\mathrm{keV}}, \label{AmatiEq}
\end{equation}
where the two calibration constants $a, b$ are Amati coefficients. The bolometric isotropic-equivalent radiated energy $E_{\rm iso}$ is converted from the observable bolometric fluence $S_{\rm bolo}$ via
\begin{equation}
  E_{\rm iso} = \frac{4\pi d_L^2 S_{\rm bolo}}{1+z}, \label{Eiso}
\end{equation}
where $d_L$ is the luminosity distance to the source. Equations~(\ref{AmatiEq}-\ref{Eiso}) link the cosmological dependent $d_L$ to GRB observables, with uncertainties in two folds. One source of the uncertainties arises from selection and instrumental effects, which has been widely investigated in Refs.~\cite{Amati:2006ky,Ghirlanda:2006ax,Ghirlanda_2008,Butler_2009,Nava,Amati:2013sca,Heussaff_2013,mochkovitch_2015,Demianski:2016zxi,Dainotti_2017}. A more challenging issue is the circularity problem, which questions whether the Amati relation calibrated in a particular cosmology can be used to distinguish cosmological models{~\cite{Kodama:2008dq, Amati:2018tso}.}

To avoid the circularity problem, Ref.~\cite{Lin:2015kaa} used cosmologies calibrated with Type Ia supernova to investigate the GRB data. The authors considered three models: $\Lambda$CDM in which the dark energy is a cosmological constant, $w$CDM where dark energy is treated as a perfect fluid with a constant equation of state $w$, and the $w_0$-$w_a$CDM model that parameterizes the dark energy equation of state as a linear function of the scale factor: $w=w_0+w_a\frac{z}{1+z}$~\cite{Chevallier:2000qy,Linder:2002et}. For all the three models, whose parameters are constrained by supernova data, the authors found that the Amati coefficients calibrated by low-redshift ($z<1.4$) GRB data is in more than $3\sigma$ tension with those calibrated by high-redshift ($z>1.4$) GRB data.

The result of Ref.~\cite{Lin:2015kaa} relies on supernova data and particular assumptions about dark energy. It is unclear whether the  $\gtrsim 3\sigma$ tension between low- and high-redshift Amati coefficients indicates a problem of Amati relation, or inconsistency between supernovae and GRB data, or failure of the dark energy models. Direct investigation by~Ref.~\cite{Wang:2017lng}, using only GRB data, found a similar tension between Amati coefficients at low and high redshifts. However, because Ref.~\cite{Wang:2017lng} assumes $\Lambda$CDM cosmology, the authors could not rule out the possibility that the tension is caused by a wrong cosmology.

To clarify all these problems, we study in this work how cosmology and selection bias play roles in the tension between low- and high-redshift Amati coefficients. We extend the GRB data set by including more samples from recent publications~\cite{Liu:2014vda,Wang:2015cya}, and use the Parameterization based on the cosmic Age (PAge) to cover a broad class of cosmological models. PAge, which will be introduced below in details, is an almost model-independent scheme recently proposed by~Ref.~\cite{Huang:2020mub} to describe the background expansion history of the universe.

\section{PAge Approximation}

PAge uses three dimensionless parameters to describe the late-time expansion history of the universe. The reduced Hubble constant $h$ measures the current expansion rate of the universe $H_0 = 100 h\hunit$. The age parameter $\page \equiv H_0t_0$ measures the cosmic age $t_0$ in unit of $H_0^{-1}$. The $\eta$ parameter characterizes the deviation from Einstein de-Sitter universe (flat CDM model), which in PAge language corresponds to $\page=\frac{2}{3}$ and $\eta=0$. The standard flat $\Lambda$CDM model with matter fraction $\Omega_m$, for instance, can be well approximated by
\begin{eqnarray}
  \page &=& \frac{2}{3\sqrt{1-\Omega_m}}\ln\frac{1+\sqrt{1-\Omega_m}}{\sqrt{\Omega_m}}; \label{lcdm1}\\
  \eta &=& 1-\frac{9}{4}\Omega_m\page^2. \label{lcdm2}
\end{eqnarray}

PAge models the Hubble expansion rate $H=-\frac{1}{1+z}\frac{dz}{dt}$, where $t$ is the cosmological time, as
\be
\frac{H}{H_0} = 1+\frac{2}{3}\left(1-\eta\frac{H_0 t}{\page}\right)\left(\frac{1}{H_0 t}-\frac{1}{\page}\right). \label{eq:page}
\ee
This equation with $\eta < 1$, which we always enforce in PAge, guarantees the following physical conditions.
\begin{enumerate}
\item{At high redshift $z\gg 1$, the expansion of the universe has an asymptotic matter-dominated $\frac{1}{1+z}\propto t^{2/3}$ behavior. (The very short radiation-dominated era is ignored in PAge.) }
\item{Luminosity distance $d_L$ and comoving angular diameter distance $d_c$ are both monotonically increasing functions of redshift $z$.}
\item{The total energy density of the universe is a monotonically decreasing function of time ($dH/dt < 0$).}
\end{enumerate}
A recent work~\cite{Luo:2020ufj} shows that, for most of the physical models in the literature, PAge can approximate the luminosity distances $d_L(z)$ ($0<z<2.5$) to subpercent level. We extend the redshift range to $z\sim 10$ and find PAge remains to be a good approximation. This is because most physical models do asymptotically approach the $\frac{1}{1+z}\propto t^{2/3}$ limit at high redshift, in accordance with PAge. {A few typical examples comparing concrete models and their PAge approximation are shown in Figure~\ref{fig:DL}.}

\begin{figure}
\includegraphics[width=\figwidth]{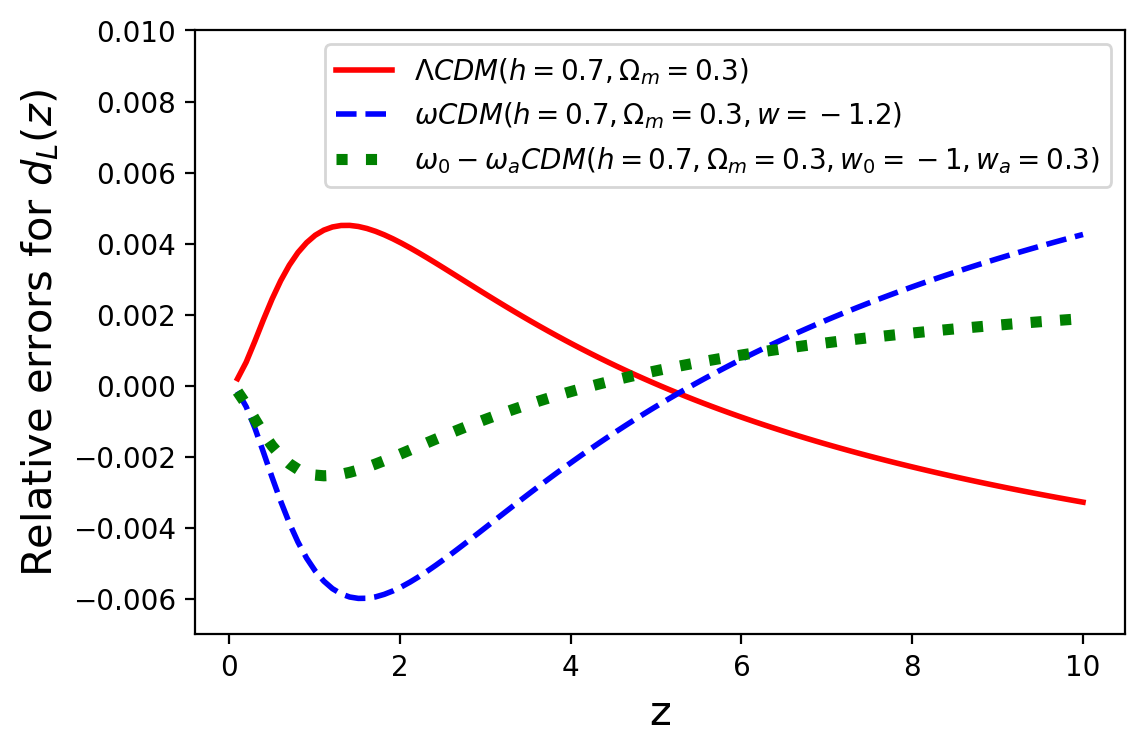}
\caption{{Relative errors of $d_L$ between $\Lambda$CDM, constant-$w$, and $w_0$-$w_a$ model and their PAge approximations, respectively. The PAge parameters are determined by matching Hubble constant and deceleration parameter at redshift zero. \label{fig:DL}}}
\label{fig:DL}
\end{figure}

Throughout this work we assume a spatially flat universe, which is well motivated by inflation models and observational constraints from cosmic microwave background~\cite{Aghanim:2018eyx}.

\section{GRB data}

\begin{figure}
\includegraphics[width=\figwidth]{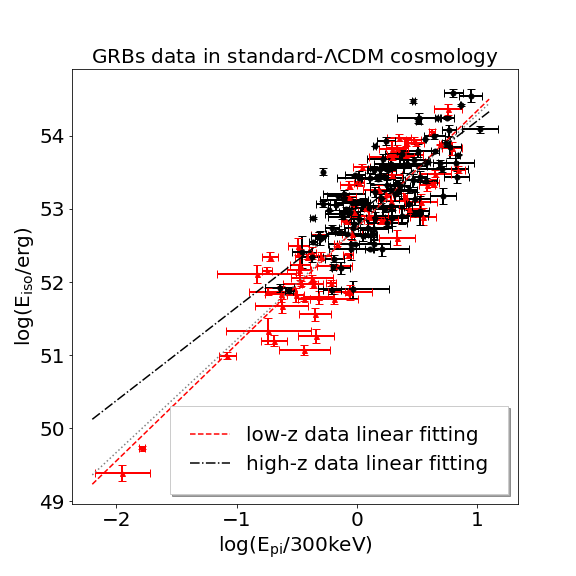}
\caption{Amati relation for low- and high-redshift data groups, respectively. A flat $\Lambda$CDM model with $\Omega_m=0.3$ and $h=0.7$ is assumed to convert $S_{\rm bolo}$ to $E_{\rm iso}$. The red triangles with error bars are 70 low-$z$ ($z\le 1.4$) GRB data in our sample. The red dashed line is their least square fitting with intercept $a=52.75$ and slope $b=1.60$. The black dots with error bars are 109 high-$z$ ($z>1.4$) GRBs data in our sample. The black dot-dashed line is their least square fitting with intercept $a=52.93$ and slope $b=1.27$. The gray dotted line shows how the low-$z$ fitting varies when the two ``outliers'' with $E_{\rm iso}<10^{50}\mathrm{erg}$ are removed from low-$z$ group. \label{fig:data}}
\end{figure}

We construct our data samples  by collecting 138 GRBs from~Ref.~\cite{Liu:2014vda} and 42 GRBs from Ref.~\cite{Wang:2015cya}. We find that GRB100728B appears in both data groups, and use weighted average algorithm to combine the two data points. Following~Ref.~\cite{Lin:2015kaa} we use $z=1.4$ to split the low-$z$ and high-$z$ samples, whose Amati relations are shown in Figure~\ref{fig:data} with red dashed line and black dot-dashed line, respectively.  For better visualization we used a fixed $\Lambda$CDM cosmology with $\Omega_m = 0.3$ and $h=0.7$. It is almost visibly clear that the low-$z$ and high-$z$ fittings of Amati relation have a discrepancy for the fixed cosmology. {We also show that the discrepancy is not dominated by the two ``outliers'' with very low $E_{\rm iso}<10^{50}\mathrm{erg}$, by removing them in the low-$z$ sample and re-plotting the linear fitting with the dotted gray line.}

To quantify this low-$z$ and high-$z$ discrepancy of Amati coefficients, and to take into account the variability of cosmologies, we now proceed to describe the joint likelihood.

For a GRB sample with
\be
x:=\log_{10} \frac{E_p}{300\mathrm{keV}},\ \ y := \log_{10} \frac{ \frac{4\pi d_L^2 S_{\rm bolo}}{1+z}, \label{xydef}}{\mathrm{erg}}
\ee
and uncertainties
\be
\sigma_x=\frac{\sigma_{E_p}}{E_p\ln 10},\sigma_y=\frac{\sigma_{S_{\rm bolo}} }{S_{\rm bolo}\ln 10},
\ee
its likelihood reads
\be
\mathcal{L}\propto \frac{ e^{-\frac{\left(y-a-bx\right)^{2}}{2\left(\sigma _{\rm int}^{2}+\sigma _y^{2}+b^{2}\sigma _x^{2}\right)}}}{\sqrt{\sigma _{\rm int}^{2}+\sigma _y^{2}+b^{2}\sigma _x^{2}}},
\ee
where  $\sigma _{\rm int}$  is an intrinsic scatter parameter representing uncounted extra variabilities~\cite{DAgostini:2005mth}. The full likelihood is the product of the likelihoods of all GRB samples in the data set. It depends on five parameters $\page$, $\eta$, $\aeff$, $b$, and $\sigma_{\rm int}$. The dimensionless Hubble parameter $h$ is absorbed into the Amati coefficient $a$ for apparent degeneracy.

We adopt a Python module {\bf emcee} \cite{ForemanMackey2013emceeTM} to perform Monte Carlo Markov Chain (MCMC) analysis for the calibration. Uniform priors are applied on  $p_{\rm age}\in[0.84,1.5]$, $\eta \in [-1,1]$, $\aeff\in[52,54]$, $b\in[-1,2]$ and $\sigma_{\rm int}\in[0.2,0.5]$.

\section{Analysis}\label{results}

\begin{table*}
\caption{\label{table:calibration} {Marginalized constraints (mean and  68\% confidence level bounds) on parameters, for all GRB samples without debiasing}}\centering
\begin{tabular}{llllll}
\hline\hline
Samples & \qquad$\page$ & \qquad $\eta$ & $a+2\log_{10}\frac{h}{0.7}$ & \qquad$b$ & \qquad$\sigma_{\rm int}$ \\
\hline
low-$z$  GRBs&$1.13^{+0.10}_{-0.27} $&$-0.08^{+0.68}_{-0.88}$ & $52.79^{+0.09}_{-0.08}$&$1.64\pm 0.10$&$0.42\pm 0.04$ \\
high-$z$ GRBs&$1.14^{+0.12}_{-0.24} $&$-0.01\pm 0.57$&$53.04^{+0.14}_{-0.12}$&$1.23 \pm 0.10 $ & $0.35\pm 0.03$ \\
low-$E_{\rm iso}$  GRBs&$0.95^{+0.02}_{-0.10} $&$-0.14^{+0.52}_{-0.83}$ & $52.57^{+0.07}_{-0.09}$&$1.28\pm 0.10$&$0.34\pm 0.03$ \\
high-$E_{\rm iso}$ GRBs&$1.11^{+0.10}_{-0.20}$&$-0.03\pm 0.57$&$53.35\pm 0.11$&$0.76 \pm 0.11 $ & $0.30^{+0.02}_{-0.03}$ \\
all GRBs &$0.93^{+0.02}_{-0.08}$&$0.04^{+0.83}_{-0.64} $ & $52.80^{+0.06}_{-0.07}$ & $1.49\pm 0.07$ & $0.38^{+0.02}_{-0.03}$ \\
\hline
\end{tabular}
\end{table*}

\begin{figure*}
\includegraphics[width=\figwidth]{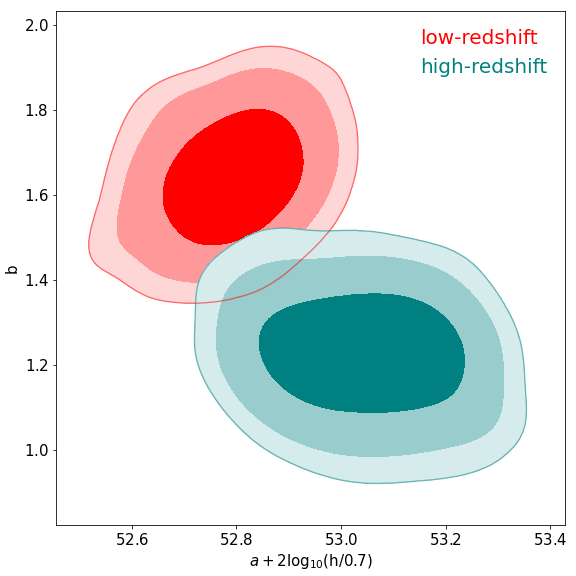}\includegraphics[width=\figwidth]{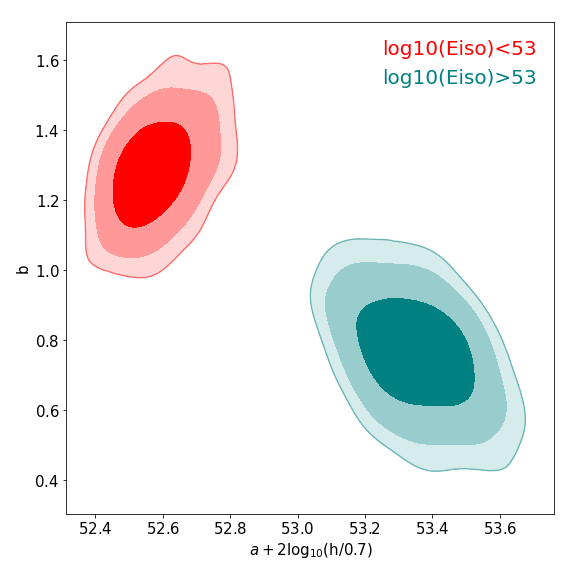}
\caption{Marginalized $1\sigma$, $2\sigma$ and $3\sigma$ contours for the Amati coefficients $a$ and $b$ for GRB data {\it without debiasing}. The left panel compares the low-$z$ ($z<1.4$) and high-$z$ ($z>1.4$) data groups. The right panel compares the low-$E_{\rm iso}$ ($E_{\rm iso}<10^{53}\mathrm{erg}$) and high-$E_{\rm iso}$ ($E_{\rm iso}>10^{53}\mathrm{erg}$) data groups.}
\label{contour}
\end{figure*}

In Table~\ref{table:calibration}, we present the posterior mean and {68\% confidence level bounds} of PAge parameters, Amati coefficients, and the intrinsic scatter $\sigma_{\rm int}$. The marginalized posteriors on PAge parameters $\page$ and $\eta$ are consistent with $\Lambda$CDM model with $\Omega_m\sim 0.3$, i.e., $(\page \sim 0.96, \eta\sim 0.37)$ as given by Eqs.~(\ref{lcdm1}-\ref{lcdm2}). {The left panel of Figure~\ref{contour} shows the marginalized $1\sigma$, $2\sigma$ and $3\sigma$ contours for the Amati coefficents  $(\aeff,b)$. The $3.0\sigma$ tension between low-$z$ and high-$z$ Amati coefficients, as previously found in Ref.~\cite{Wang:2017lng} for $\Lambda$CDM model, persists here for the much less model-dependent PAge cosmology.}

The result seems to suggest redshift evolution of Amati relation. However, it does not mean that cosmological environment can have an impact on GRB physics. A more likely explanation is that the high-redshift samples are biased samples due to flux limits in observations. To test this conjecture, we re-group the samples by $E_{\rm iso}$ rather than by redshift~\footnote{We use the best-fit values of $p_{\rm age}$ and $\eta$ calibrated by all GRBs samples to calculate the luminosity distance $d_L$ that is needed to compute $E_{\rm iso}$.}. More specifically, we choose a roughly median value $10^{53}\mathrm{erg}$ as the splitting boundary to divide the data samples into low-$E_{\rm iso}$ and high-$E_{\rm iso}$ groups. The MCMC results for the two data groups are shown in Table~\ref{table:calibration}. The marginalized posteriors of $(\aeff,b)$ are shown the right panel of Figure~\ref{contour}, from which we find a much stronger $\sim 6 \sigma$ tension between low-$E_{\rm iso}$ and high-$E_{\rm iso}$ Amati coefficients. This result suggests that {Amati's linear relation is not universal across all energy ($E_{\rm iso}$) scales, and the seemingly redshift evolution of Amati coefficients may just be a selection effect due to flux limits in observations, which is often referred to as Malmquist bias in the astronomy community.}

\begin{table*}
\caption{\label{table:debiasdata} {Marginalized constraints (mean and  68\% confidence level bounds) on parameters, for the debiased ($E_{\rm iso}$ distribution matched) GRB samples}}\centering
\begin{tabular}{llllll}
\hline\hline
Samples & $\qquad \page$ & $\qquad \eta$ & $a+2\log_{10}\frac{h}{0.7}$ & $\qquad b$ & $\qquad \sigma_{\rm int}$\\
\hline
low-$z$  GRBs&$1.13^{+0.11}_{-0.26} $&$-0.12^{+0.49}_{-0.83}$ & $52.90\pm 0.09$&$1.48\pm 0.13$&$0.39^{+0.04}_{-0.05}$\\
high-$z$ GRBs&$1.14^{+0.13}_{-0.26} $&$0.08\pm 0.58$&$52.94^{+0.14}_{-0.12}$&$1.35 \pm 0.14 $ & $0.31^{+0.03}_{-0.04}$\\
low-$E_{\rm iso}$ &$1.02^{+0.06}_{-0.17}$&$-0.29^{+0.24}_{-0.70} $ & $52.67^{+0.08}_{-0.11}$ & $1.13\pm 0.12$ & $0.30^{+0.03}_{-0.04}$\\
high-$E_{\rm iso}$ &$1.14^{+0.12}_{-0.23}$&$0.05^{+0.88}_{-0.75} $ & $53.33^{+0.13}_{-0.12}$ & $0.86\pm 0.19$ & $0.31^{+0.03}_{-0.05}$\\
\hline
\end{tabular}
\end{table*}

\begin{figure}
\includegraphics[width=\figwidth]{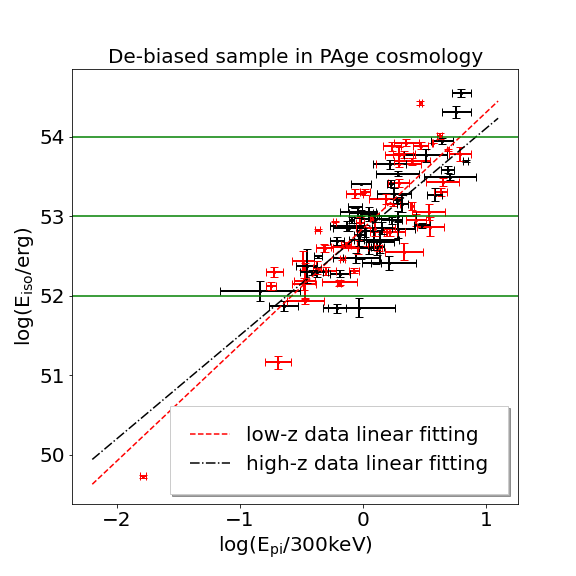}
\caption{Amati relation for {debiased ($E_{\rm iso}$-distribution matched)} low- and high-redshift groups, respectively. The debiasing is done by matching number of GRB samples in each $E_{\rm iso}$ bin separated by green lines. A best-fit flat PAge model with $p_{\rm age}=0.93$ and $\eta=0.04$ is used to convert $S_{\rm bolo}$ to $E_{\rm iso}$. The red dashed line is least square fitting of low-z data with intercept $a=52.84$ and slope $b=1.46$. The black dot-dashed line is least square fitting of high-z data with intercept $a=52.80$ and slope $b=1.30$.\label{fig:matched}}
\label{fig:debias}
\end{figure}

\begin{figure*}
\includegraphics[width=\figwidth]{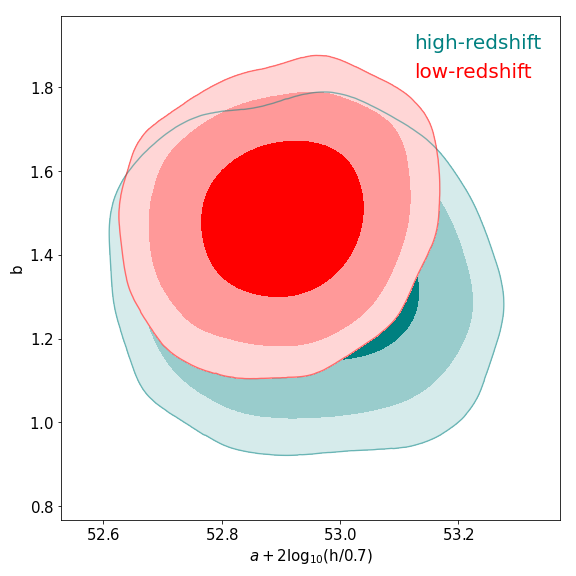}\includegraphics[width=\figwidth]{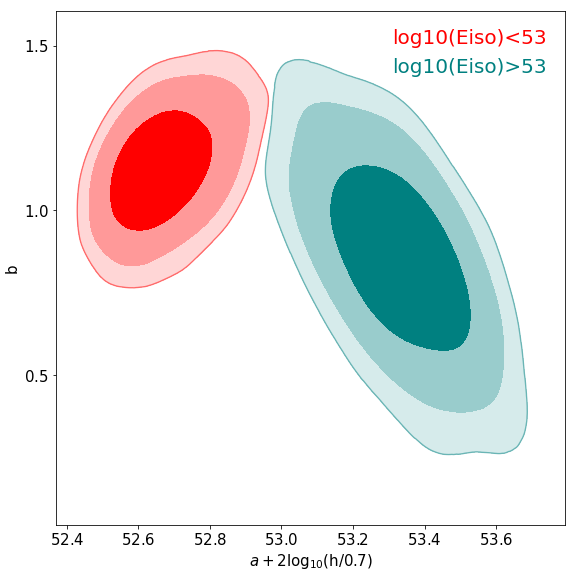}
\caption{Marginalized $1\sigma$, $2\sigma$ and $3\sigma$ contours for the Amati coefficients $a$ and $b$ for {\it debiased ($E_{\rm iso}$-distribution matched)} GRB samples. The left panel compares the low-$z$ ($z<1.4$) and high-$z$ ($z>1.4$) data groups. The right panel compares the low-$E_{\rm iso}$ ($E_{\rm iso}<10^{53}\mathrm{erg}$) and high-$E_{\rm iso}$ ($E_{\rm iso}>10^{53}\mathrm{erg}$) data groups.}
\label{low-highpart}
\end{figure*}

To further test the $E_{\rm iso}$ selection bias, we split the data into four $E_{\rm iso}$ bins, with $E_{\rm iso}<10^{52}\mathrm{erg}$, $10^{52}\mathrm{erg}\le E_{\rm iso}<10^{53}\mathrm{erg}$ , $10^{53}\mathrm{erg}\le E_{\rm iso}<10^{54}\mathrm{erg}$ , and $E_{\rm iso}\ge 10^{54}\mathrm{erg}$, respectively. For GRBs in each $E_{\rm iso}$ bin, we {sort the GRB redshifts and discard samples, as uniformly in redshift as possible, from either the low-$z$ data group or the high-$z$ data group (whichever contains more GRB samples), such that the numbers of samples in two groups coincide.} After {this process}, the low-$z$ and high-$z$ groups have roughly the same $E_{\rm iso}$ distribution, and thus should be debiased. {About half of the GRB samples are discarded during this debiasing process.} The {remaining debiased ($E_{\rm iso}$-distribution matched)} samples and their low-$z$ and high-$z$  Amati coefficients are shown in Figure~\ref{fig:matched}.  The marginalized posteriors of Amati coefficients for the debiased low-$z$ and high-$z$ samples, as shown in {Table~\ref{table:debiasdata} and} the left panel of Figure~\ref{low-highpart}, suggest no redshift evolution of Amati relation at all. {For a comparison, we also split the debiased samples into low-$E_{\rm iso}$ ($E_{\rm iso}<10^{53}\mathrm{erg}$) and high-$E_{\rm iso}$ ($E_{\rm iso}>10^{53}\mathrm{erg}$) data groups and repeat the analysis. The results are shown in Table~\ref{table:debiasdata} and the right panel of Figure~\ref{low-highpart}. In this case we find a $3.8\sigma$ tension in Amati coefficients. In summary, after the debiasing process, the redshift evolution of Amati coefficients vanishes, while the evidence for $E_{\rm iso}$-dependence of Amati relation remains to be strong.}

Finally, we upload chains and analysis tools to \url{https://zenodo.org/record/4600891} to allow future researchers to reproduce our results.

\section{Conclusions and Discussion}

By marginalizing over PAge parameters, we have effectively studied Amati relation in a very broad class of cosmologies, and to certain extent, have avoided the circularity problem. Besides the PAge framework, phenomenological extrapolation methods such as Taylor expansion, Pad\'e approximations, and Gaussian process, can also be used to calibrate GRB luminosity correlations~\cite{Lusso:2019akb, Mehrabi:2020zau, Luongo:2020aqw, Luongo:2020hyk}. These models typically contain many degrees of freedom and are poorly constrained with current GRB data. With future increasing number of GRBs, it would be interesting to compare them with PAge approach.

The $\sim 3\sigma$ tension between low-redshift and high-redshift Amati coefficients,  previously found in Ref.~\cite{Wang:2017lng} for $\Lambda$CDM, {persists} for the broad class of models covered by PAge. The insensitivity to cosmology of the redshift evolution of GRB luminosity correlation may indicate that the redshift evolution is partially trackable without assuming a cosmology. A straightforward method, as is done in Ref.~\cite{Wang:2017lng}, is to treat the Amati coefficients $a$ and $b$ as some functions of redshift. The disadvantage is that the choice of functions $a(z)$ and $b(z)$ is somewhat arbitrary, and their parameterization may introduce too many degrees of freedom. From observational perspective, Ref.~\cite{Izzo:2015vya} proposes to replace Amati relation with a more complicated {\it Combo} correlation, which uses additional observables from the X-ray afterglow light curve. This approach seems to be more competitive and is now widely studied in the literature~\cite{Luongo:2020aqw, Luongo:2020hyk}.

Our further investigation reveals that Amati relation is indeed non-universal, and strongly depends on the energy scale ($E_{\rm iso}$ range). {The $E_{\rm iso}$ dependence of Amati relation differs from the circularity problem. It calls for a better modeling of GRB luminosity correlation, rather than just focusing on  how to reduce the impact of the fiducial cosmology. By matching the low-$z$ and high-$z$ $E_{\rm iso}$ distributions, we confirm that the discrepancy between low-$z$ and high-$z$ Amati coefficients can be fully interpreted as a selection effect (Malmquist bias), and does {\it not} imply cosmological evolution of GRBs.}

{Although the $E_{\rm iso}$-dependence of Amati relation makes GRB a more sophisticated (an arguably, worse) quasi-standard candle, precision cosmology with GRBs remains to be an exciting possibility. While the statistical power of GRB data will foreseeably increase, subtle issues may also arise in the future. For instance, while many GRB samples at different redshift and observed in different energy bands are combined to achieve better statistics, photometric calibration, which currently has become a major limitation of SN cosmology~\cite{Kenworthy21, Pierel:2020udq}, may also be an issue that deserves some investigation in GRB cosmology.}

\section{Acknowledgements}

{We are grateful to the anonymous referees who made many insightful comments that are very helpful for improving the quality of this work.} This work is supported by the National key R\&D Program of China (Grant No. 2020YFC2201600), National SKA Program of China No. 2020SKA0110402, Guangdong Major Project of Basic and Applied Basic Research (Grant No. 2019B030302001), and National Natural Science Foundation of China (NSFC) under Grant No. 12073088. 

%

\end{document}